\documentclass[aps,twocolumn,tightenlines,nofootinbib,superscriptaddress]{revtex4-1}
\usepackage[utf8]{inputenc}
\usepackage{graphicx}
\usepackage{subcaption}
\usepackage{mwe}
\usepackage{mathrsfs}
\usepackage{amsfonts}
\usepackage{setspace}
\usepackage{cellspace}
\usepackage{amsmath,amssymb,bm}
\usepackage[colorlinks=true,linkcolor=blue]{hyperref}
\usepackage{xcolor}
\usepackage{epsfig}
\usepackage{slashed}
\usepackage{caption}
\usepackage{hhline,multirow,tabularx}  
\usepackage{dcolumn}    
\usepackage{url}        
\usepackage{braket}     
\renewcommand{\bra}[1]{\left<#1\left|}
\renewcommand{\ket}[1]{\right|#1\right>}

\DeclareMathAlphabet{\mathdutchcal}{U}{dutchcal}{m}{n}
\SetMathAlphabet{\mathdutchcal}{bold}{U}{dutchcal}{b}{n}
\DeclareMathAlphabet{\mathdutchbcal}{U}{dutchcal}{b}{n}

\begin{document}

\title{Explore the Nucleon Tomography through Di-hadron Correlation in Opposite Hemisphere in Deep Inelastic Scattering}

\author{Yuxun Guo}
\email{yuxunguo@lbl.gov}
\affiliation{Nuclear Science Division, Lawrence Berkeley National
Laboratory, Berkeley, CA 94720, USA}
\author{Feng Yuan}
\email{fyuan@lbl.gov}
\affiliation{Nuclear Science Division, Lawrence Berkeley National
Laboratory, Berkeley, CA 94720, USA}

\begin{abstract}

We investigate the correlation of di-hadron productions between the current fragmentation region (CFR) and target fragmentation region (TFR) in deep inelastic scattering as a probe of the nucleon tomography. The QCD factorization and powering counting method are applied to compute the relevant diffractive parton distribution functions in the valence region. In particular, we show that the final state interaction effects lead to a nonzero longitudinal polarized quark distribution associated with the unpolarized nucleon target. 
This explains the observed beam single spin asymmetry (BSA) from a recent Jefferson Lab experiment.  
We further show that the BSA in the single diffractive hadron productions in the TFR, although kinematically suppressed, also 
exists because of the final state interaction effects.

\end{abstract}

\maketitle

\section{Introduction}

One of the most fundamental and compelling topics in hadronic physics has been the nucleon tomography imaging 
encoded in the multidimensional parton distributions~\cite{Boer:2011fh, AbelleiraFernandez:2012cc, Accardi:2012qut,AbdulKhalek:2021gbh,Gross:2022hyw,Achenbach:2023pba}.
Beyond the collinear parton distribution functions (PDFs), they may contain extra information either in the momentum space, corresponding to the transverse-momentum-dependent PDFs (TMDs)~\cite{Collins:1981uk,Collins:1981uw,Collins:1989gx,Boussarie:2023izj}, or in the coordinate space, corresponding to the generalized parton distributions (GPDs)~\cite{Muller:1994ses, Ji:1996ek, Ji:1998pc,Radyushkin:1997ki}.
The two distributions can be integrated under the even higher-dimensional phase-space distributions, the Wigner distribution~\cite{Wigner:1932eb}, which are also known as the generalized TMDs (GTMDs) in the momentum space~\cite{Ji:2003ak,Belitsky:2003nz,Lorce:2011kd}. 

\begin{figure}[ht]
    \centering
    \includegraphics[width=0.48\textwidth]{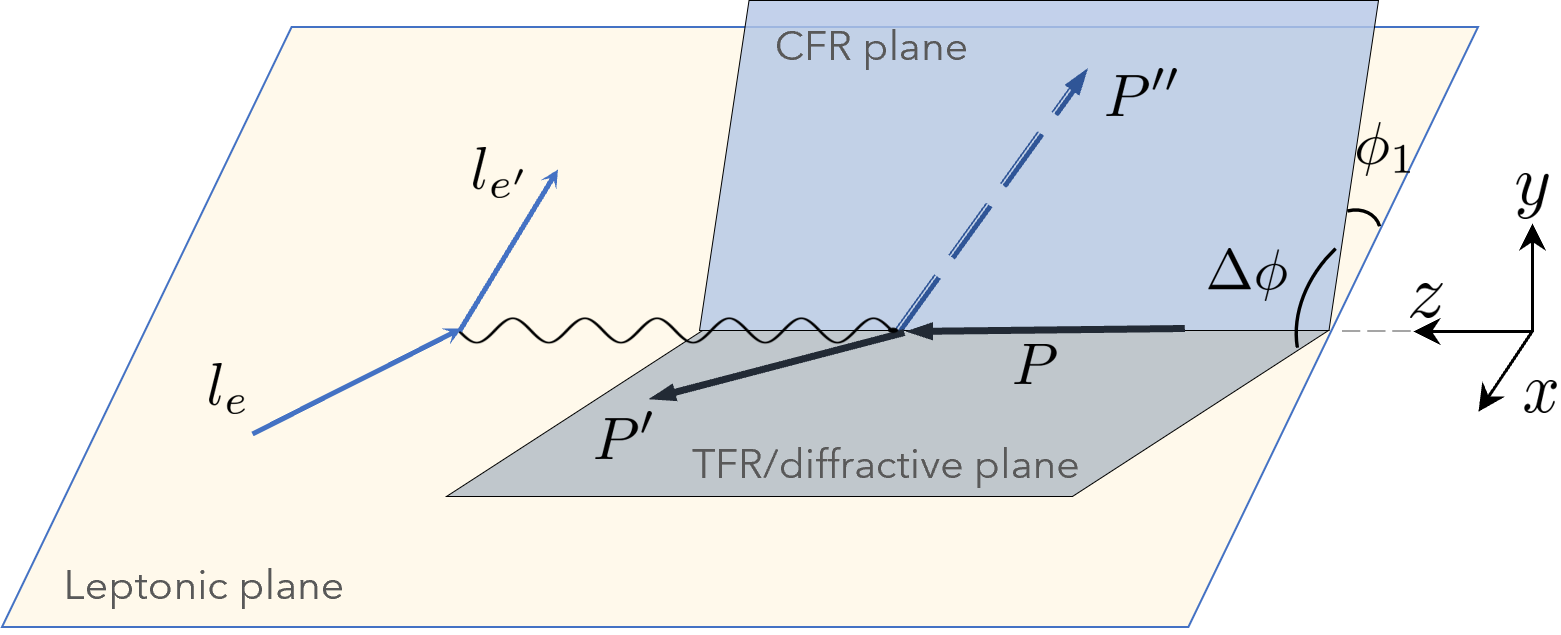}
    \caption{\raggedright An illustration of the correlated di-hadron production in the CFR and TFR. The proton in the TFR is diffractive with a low momentum transfer.}
    \label{fig:dihadronplt}
\end{figure}

Meanwhile, theoretical efforts of last few years have made progress to explore the nucleon tomography in terms of GTMDs from various experiments at the current and future facilities~\cite{Hatta:2016dxp,Ji:2016jgn,Hatta:2016aoc,Boussarie:2018zwg,Bhattacharya:2017bvs,Bhattacharya:2022vvo,Echevarria:2022ztg}.
However, the challenge remains to study them in full kinematics. Therefore, a comprehensive program to tackle this issue from all possible methods is highly recommended.
In this paper, we study the di-hadron correlations in opposite hemisphere along the beam direction in deep inelastic scattering (DIS) as a unique probe to the nucleon tomography.
In the usual DIS kinematics, as shown in Fig.~\ref{fig:dihadronplt}, this corresponds to the correlation between hadron productions in the current fragmentation region (CFR) and target fragmentation region (TFR)~\cite{Anselmino:2011ss,Anselmino:2011bb,Boglione:2016bph,Chen:2021vby,Chen:2023wsi}. 

Different from the semi-inclusive DIS (SIDIS) in the CFR where factorization in terms of TMDs and fragmentation functions (FFs) has been thoroughly studied~\cite{Collins:1981uw,Collins:1981uk,Collins:1989gx,Ji:2004wu,Ji:2004xq,Boussarie:2023izj}, the SIDIS in the TFR has not been extensively explored in the literature, albeit the factorization theorem in terms of the so-called fracture functions~\cite{Trentadue:1993ka,Berera:1995fj,Grazzini:1997ih} has been shown~\cite{Graudenz:1994dq,deFlorian:1995fd,Collins:1997sr} and efforts have been made to understand the fracture functions in different kinematic regions~\cite{Anselmino:2011ss,Anselmino:2011bb,Chai:2019ykk,Chen:2021vby}. Based on these developments, we will investigate the di-hadron correlation between current and target fragmentation regions, focusing on the kinematics of proton (or other baryon) diffractive in the TFR. We refer to this process as the semi-inclusive diffractive DIS (SIDDIS) in this paper. It is interesting to note that a nontrivial beam spin asymmetry (BSA) in correlated di-hadron productions in SIDDIS ($ep \to e p' \pi^+ X$) has been observed in experiments at Jefferson Lab (JLab)~\cite{CLAS:2022sqt}. The goal of the current paper is to investigate this physics and demonstrate the potential of probing nucleon tomography from future experiments at JLab and electron-ion collider (EIC)~\cite{Boer:2011fh, AbelleiraFernandez:2012cc, Accardi:2012qut,AbdulKhalek:2021gbh,Gross:2022hyw,Achenbach:2023pba}.

Meanwhile, at small-$x$, which is relevant for EIC experiments, the current knowledge of similar type fracture functions comes from diffractive DIS measurements at HERA~\cite{H1:1994ahk,H1:1995cha,H1:1997bdi,H1:2006zyl,H1:2012pbl}, 
where they are also called as the diffractive PDFs (DPDFs)~\cite{Collins:1997sr,deFlorian:1998rj,Goharipour:2018yov,Salajegheh:2022vyv,Salajegheh:2023jgi,Armesto:2019gxy}.  
Studies in the Color-Glass Condensate (CGC) formalism have established interesting connections of the dipole S-matrices, which is equivalent to gluon Wigner distribution at small-$x$~\cite{Hatta:2016dxp}, with the DPDFs in SIDDIS~\cite{Hebecker:1997gp,Buchmuller:1998jv,Golec-Biernat:1999qor,Hautmann:1999ui,Hautmann:2000pw,Golec-Biernat:2001gyl,Iancu:2021rup,Hatta:2022lzj,Iancu:2022lcw}. 
In the following, we will focus on the moderate-$x$ range and argue that the same connection exists for the quark sector. 
While the fracture functions cannot be calculated with perturbative quantum chromodynamics (QCD) in general, the target fragmentation process will be approximated with the hadronization of one knocked-out parton, whereas the diffractive nucleon results in certain exclusive matrix elements. We argue that a factorization theorem of such semi-inclusive processes in terms of the exclusive matrix elements can be established following previous examples~\cite{Collins:1997sr} and this semi-exclusive picture will provide some insights on this rather unexplored phenomenon. 

We notice a related observable in DIS, called the nucleon energy-energy correlator, has also been proposed to study nucleon structure in the target fragmentation region~\cite{Liu:2022wop,Cao:2023oef,Liu:2023aqb,Li:2023gkh}. The physics discussed in these papers is substantially different, but complementary to what we focus on in this paper.  

\section{Diffractive PDFs and exclusive matrix elements}

\label{sec:dpdf}

For semi-inclusive hadron productions in DIS, we can formulate the amplitude squared  as
\begin{equation}
    \sum_{X}\left|\mathcal{M}(lP\to l' P' X)\right|^2=\frac{e^4}{q^4}l^{\mu\nu} H_{\mu\nu}\ ,
\end{equation}
in terms of the leptonic and hadronic tensors $l^{\mu\nu}$ and $H_{\mu\nu}$. The leptonic tensor $l^{\mu\nu}$ can be derived with QED, whereas the hadronic tensor $H_{\mu\nu}$ can be written with the non-perturbative nucleon matrix elements as,
\begin{align}
\begin{split}
        H_{\mu\nu}=\sum_{X}\int \frac{\text{d}^4 r}{(2\pi)^4} e^{iq\cdot r} \bra{P}J^{\mu}(r)\ket{P'X}\bra{P'X}J^{\nu}(0)\ket{P}\ .
\end{split}
\end{align}
The $X$ here refers to all unidentified final-state particles that will be summed over. The nucleon polarization vectors $S$ and $S'$ are and will be suppressed, unless specified otherwise. 
The factorization formula in terms of the fracture function has been shown as,
\begin{equation}
    H^{\mu\nu} = T^{\mu\nu}_{\alpha\beta} \mathcal {M}^{\alpha\beta}(l,P,P')\ ,
\end{equation}
with the hard-scattering amplitude $T^{\mu\nu}_{\alpha\beta}$, the fracture function $\mathcal {M}^{\alpha\beta}(l,P,P')$, and  Dirac indices $\alpha$, $\beta$. The fracture function is defined as~\cite{Berera:1995fj},
\begin{align}
\begin{split}
    \mathcal{M}^{\alpha\beta}(l,P,P'&) = \sum_{X}\int\frac{\text{d}^3P_X}{(2\pi)^3 2E_{X}} \int\frac{\text{d}^4 r}{(2\pi)^4}e^{ir\cdot l} \\
    &\bra{P} \bar \psi^{\alpha}(0)\ket{P';X}\bra{P';X} \psi^{\beta}(r) \ket{P}\ ,
\end{split}
\end{align}
with implicit gauge links between the two fields. In principle, the fracture functions/DPDFs are not calculable. However, in certain kinematics, we can apply perturbative QCD and power counting analysis to evaluate their behaviors. For example, the small-$x$ DPDFs can be calculated in terms of dipole amplitude, from which the transverse momentum dependence and distribution function behavior can be derived in a model-independent way~\cite{Hebecker:1997gp,Buchmuller:1998jv,Golec-Biernat:1999qor,Hautmann:1999ui,Hautmann:2000pw,Golec-Biernat:2001gyl,Iancu:2021rup,Hatta:2022lzj,Iancu:2022lcw}. In the following, we argue that, at large-$x$, we can apply the perturbative QCD to derive the power behaviors of the quark diffractive PDFs.

\begin{figure}[ht]
    \centering
    \begin{subfigure}{.238\textwidth}
        \centering
        \includegraphics[width=0.89\textwidth]{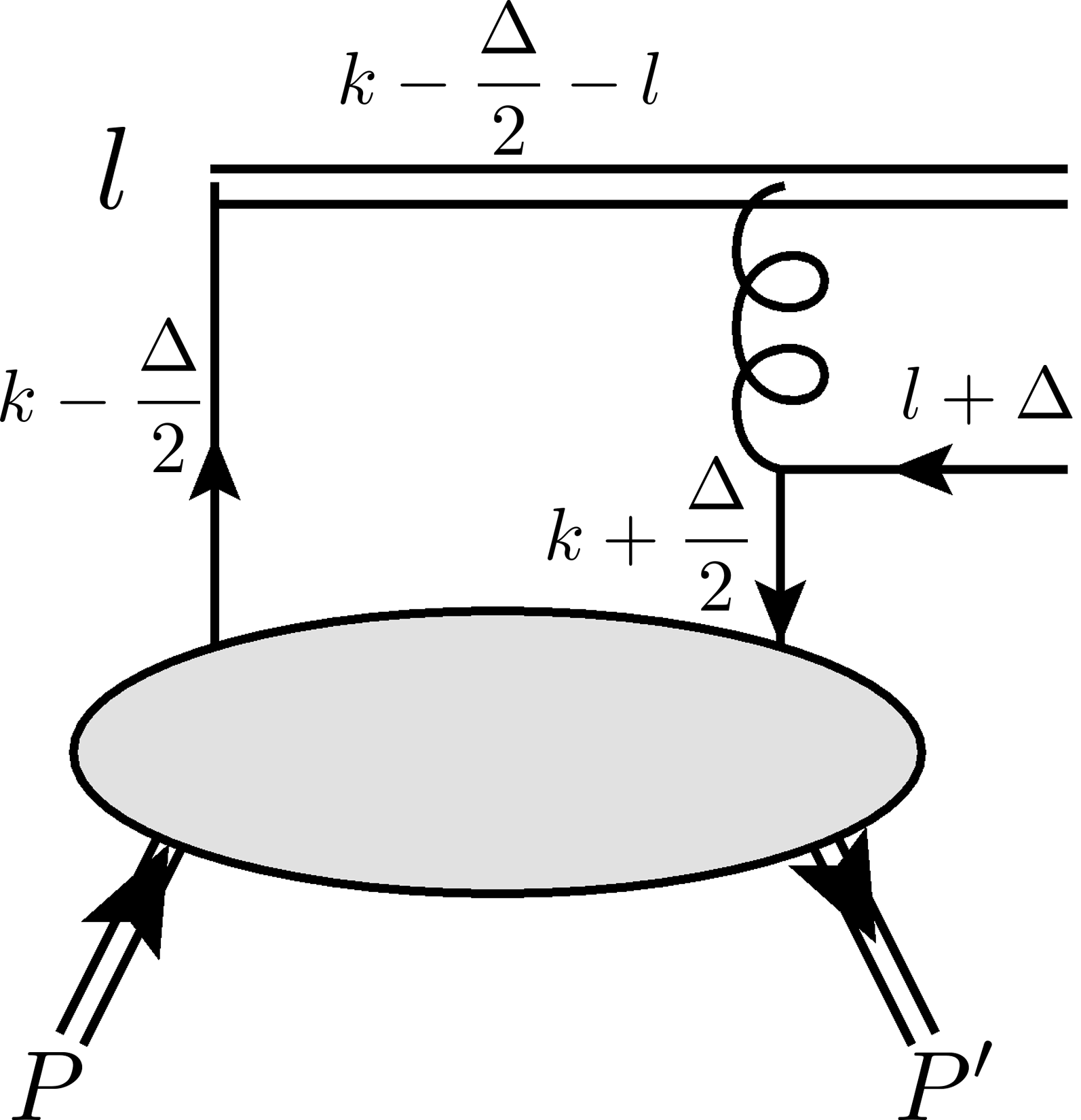}
        \caption{\label{fig:dpdffromgpda}}
    \end{subfigure} %
    \begin{subfigure}{.238\textwidth}
        \centering
        \includegraphics[width=0.9\textwidth]{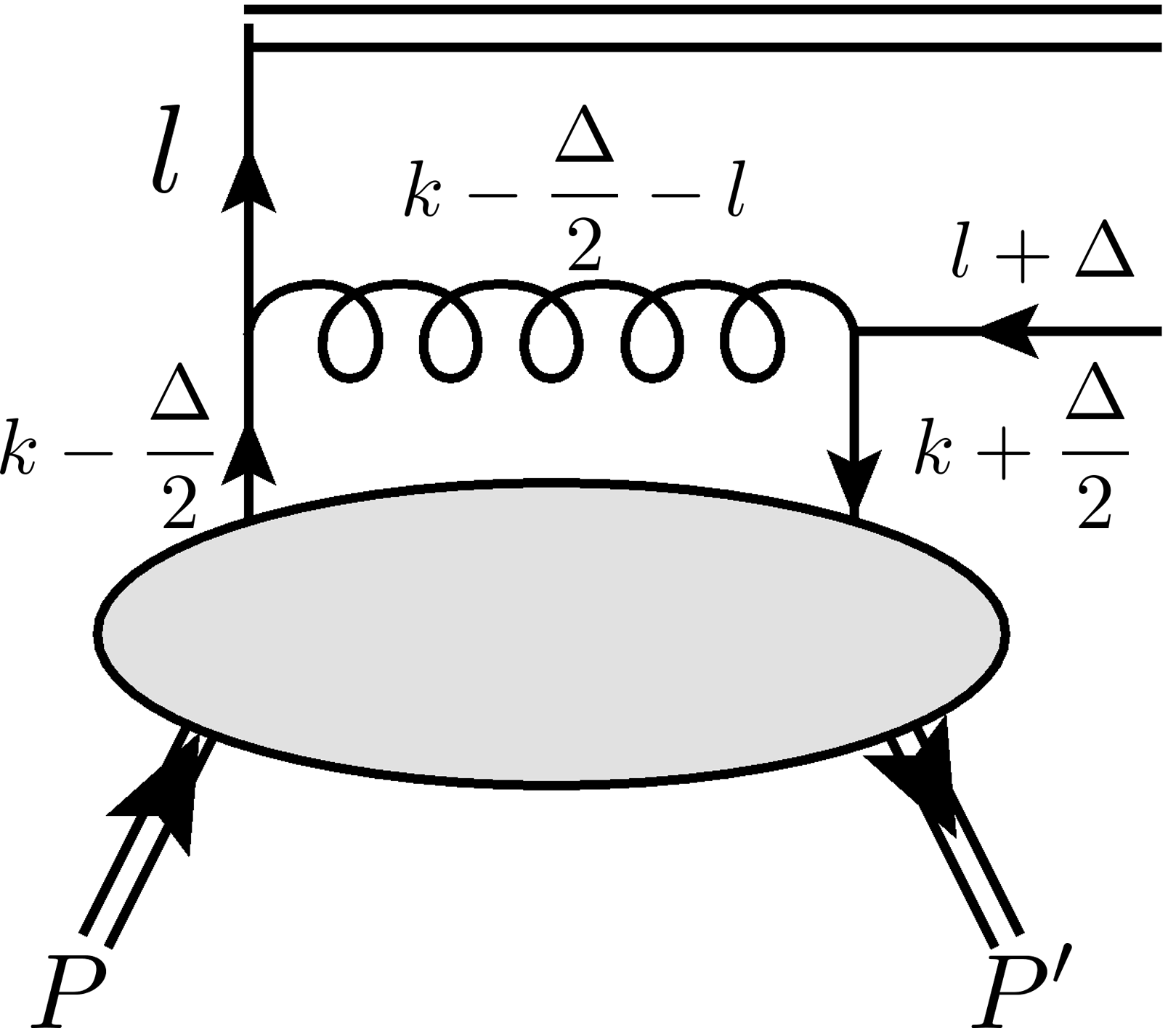}
        \caption{}
    \end{subfigure} %
    \caption{\raggedright  Leading-order Feynman diagrams of diffractive PDFs in terms of the exclusive matrix elements, such as the GTMDs or GPDs. The hadronization of the parton moving to the right not shown explicitly accounts for the target fragmentation process, and the DPDFs will be the square of these diagrams.}
    \label{fig:dpdffromgpd}
\end{figure}

The key ingredient in the perturbative analysis is to project the dependence on the nucleon target with the exclusive matrix elements of the nucleon, such as the GTMDs and GPDs. In Fig.~\ref{fig:dpdffromgpd}, we show the typical diagrams at the leading order. Gluon exchange is needed to make the semi-inclusive calculation possible. The perturbative QCD can be applied in the large-$x$ region because: (1) large transverse momentum of the struck quark probed in these quark DPDFs; (2) large-$x$ power counting arguments where higher order corrections will be power suppressed. With these arguments, we can compute the quark DPDFs in terms of quark GTMDs and GPDs. 

Then, we can write down the quark DPDFs in terms of the exclusive matrix elements as
\begin{eqnarray}
    \mathcal {M}^{\left[\Gamma \right]}(x_B,l_\perp,P,P') =\frac{1}{4\zeta_2} \int &&\frac{\text{d}l^+\text{d}l^-}{(2\pi)^3}\delta(l^+-x_B P^+)\nonumber\\
    &&\times \text{Tr}\left[\mathdutchcal{M}^\dagger  \Gamma \mathdutchcal{M}\right]\ ,
\end{eqnarray}
where $\zeta_2\equiv P'^+/P^+= (1-\xi)/(1+\xi)$, $\Gamma$ represents the Dirac matrices to project out certain spin-dependent distributions and $\mathdutchcal{M}$ for the Feynman diagram contributions from Fig.~\ref{fig:dpdffromgpd}. The $\int\text{d}l^+$ and $\int \text{d}^3P_X (2\pi)^{-3} (2E_X)^{-1}$ integral are associated with corresponding $\delta$ functions that will trivially set $P_X = -l-\Delta$ and $l^+=x_B P^+$.

To evaluate the Feynman diagrams in Fig.~\ref{fig:dpdffromgpd}, the exclusive matrix elements associated with the nucleon target can be parameterized in terms of GTMDs, which demonstrates the crucial dependence on the nucleon tomography. To highlight this dependence and simplify the derivation, as a first step, we expand the transverse momentum in the GTMDs and make connections to the GPDs. As an example, we show the 
leading twist quark DPDFs with projections of $\gamma^+$ and $\gamma^+\gamma^5$:
\begin{widetext}
\begin{eqnarray}
    \mathcal {M}^{\left[\gamma^+ \right]} &=& \frac{1}{4\zeta_2(2\pi)^3}\frac{ 4 C_F^2 g^4}{l_\perp^4} \frac{\left(l^++\Delta^+\right)^2}{\Delta^+}\int_{-1}^1\text{d}x'\text{d}x''   \left(\left|F^{\left[\gamma^+\right]}\right|^2+\left|F^{\left[\gamma^+\gamma^5\right]}\right|^2\right)\Bigg[\frac{l^++\Delta^+}{\Delta^+}\left(\text{Re}\mathcal{T}_b^2+\text{Im}\mathcal{T}_b^2\right)\nonumber\\
    &&~~~~~~~~~~~~~~~~~~~~~~~~~~~~~~~~~~~~~~+ \text{Im}\mathcal{T}_a^2 +2 \text{Im}\mathcal{T}_a\text{Im}\mathcal{T}_b\Bigg] + \text{cross terms} \ ,\label{eq:e5}\\
    \label{eq:mgammaplus5}
    \mathcal {M}^{\left[\gamma^+\gamma 5 \right]} &=& \frac{1}{4\zeta_2(2\pi)^3}\frac{8 C_F^2 g^4}{l_\perp^6} \frac{\left(l^++\Delta^+\right)^2}{\Delta^+}\int_{-1}^1\text{d}x'\text{d}x''  \text{Im}\mathcal{T}_a \text{Re}\mathcal{T}_b  \left(\left|F^{\left[\gamma^+\right]}\right|^2+\left|F^{\left[\gamma^+\gamma^5\right]}\right|^2\right) \epsilon_\perp^{l\Delta} + \text{cross terms}\ ,\label{eq:e6}
\end{eqnarray}
\end{widetext}
for unpolarized and longitudinal polarized quark distributions, respectively, where we introduce the notation $\epsilon_\perp^{l\Delta } \equiv \epsilon_\perp^{\alpha\beta}l_{\alpha} \Delta_{\beta} $ for simplicity. The $F^{[\gamma^+]}$ and $F^{[\gamma^+\gamma^5]}$ above represent the dependence on the quark GPDs~\cite{Ji:1996ek},
\begin{align}
\begin{split}
    F_q^{\left[\gamma^+\right]}&=\frac{1}{2\bar P^+}\bar u(P',S')\left[
   \gamma^+ H_q+\frac{i\sigma^{+\nu} \Delta_\nu}{2M} E_q\right] u(P,S)\ ,\nonumber
\end{split}\\
\begin{split}
    F_q^{\left[\gamma^+\gamma^5\right]}&=\frac{1}{2\bar P^+}\bar u(P',S')\left[
   \gamma^+ \gamma^5 \widetilde H_q+\frac{ \Delta^+ \gamma^5 }{2M} \widetilde E_q\right] u(P,S)\ .
\end{split}
\end{align}
The square of them can be calculated by averaging (summing) over the polarization $S$ ($S'$) of the initial (final) proton in the case of unpolarized target, and the cross terms of them vanish. The $H_q,E_q,\widetilde H_q$ and $\widetilde E_q$ are the four well-known leading-twist quark GPDs that are functions of the three standard GPD variables $(x',\xi,t)$. These scalars are the average momentum fraction of the parton $x'\equiv k^+/\bar P^+ $ defined above, the skewness parameter $\xi \equiv -\Delta^+/(2\bar P^+)$ and the momentum transfer squared $t\equiv \Delta^2$, which will be used hereafter.  The hard scattering coefficients are defined as,
\begin{align}
    \text{Im}\mathcal{T}_a(x',\xi,x_B) &\equiv \pi \delta\left(x'+\xi-(1+\xi)x_B \right)\ , \nonumber\\
    \text{Re}\mathcal{T}_b(x',\xi) &\equiv -\text{P.V.} \frac{1 }{x'-\xi}\ ,\\
    \text{Im}\mathcal{T}_b(x',\xi) &\equiv  \pi \delta\left(x'-\xi \right)\nonumber \ ,
\end{align}
that depends on $(x',\xi)$ in general. The $\text{Im}\mathcal{T}_a$ depends on $x_B$ additionally, resulting from the $x_B$-dependence of the momentum fraction $l^+ = x_B P^+$. 

It is clear that the interference between the imaginary part of amplitude ${\cal T}_a$ and the real part of amplitude ${\cal T}_b$ plays a crucial role in having a nonzero result for the projection of ${\cal M}^{[\gamma^+\gamma^5]}$, which describes the longitudinal polarized quark distribution associated with the unpolarized nucleon target. It is interesting to note that it is the same mechanism that generates a nonzero Sivers-type TMD quark/gluon distributions~\cite{Brodsky:2002ue,Collins:2002kn,Brodsky:2002cx,Ji:2002aa,Belitsky:2002sm}. 

Furthermore, in the above results, the knocked-out parton with momentum is approximately on-shell: $P_X^2 \approx 0$. Then, one has $x_B \le x_{\mathbb{P}}$ when requiring $P_X^+\ge0$, where the so-called pomeron momentum fraction is defined as
\begin{equation}
    x_{\mathbb{P}}\equiv -\Delta^+/P^+ = 2\xi/(1+\xi)\ .
\end{equation}
For fixed transverse momentum $l_\perp$, one has $l^-\to \infty$ when approaching the $x_B \to x_{\mathbb{P}}$ limit. Consequently, the propagators carrying momentum $l$ will be suppressed by $l^2 \propto l_\perp^2 x_{\mathbb{P}}/(x_{\mathbb{P}}-x_B)\to \infty$, which plays an essential role for the power counting argument of our calculations and the parton distributions at large-$x$ in general~\cite{Drell:1969km,West:1970av,Brodsky:1973kr,Farrar:1975yb}. 
This suppression has been illustrated in an early perturbative calculation of the DPDF with exclusive nucleon matrix elements in Ref.~\cite{Berera:1995fj}. It has been shown therein that the DPDF will be suppressed in the $x_B\to x_{\mathbb{P}}$ limit in the form of $(1-\beta)^p$ with $\beta\equiv x_B/x_{\mathbb{P}}$. Based on perturbative calculations, the power $p=2$ is obtained. 

\section{Correlated di-hadron productions in the CFR and TFR}

\label{sec:dihadron}

With the above quark DPDFs, we can study the correlated di-hadron productions between CFR and TFR and the associated spin asymmetries. Here, we give an example of the BSA in this process. As discussed in the previous section, this BSA results from the non-zero DPDF projection $\mathcal{M}^{\left[\gamma^+\gamma^5\right]}$ due to final state interaction effects. More specifically, the BSA of the correlated di-hadron productions can be written as~\cite{CLAS:2022sqt}:
\begin{equation}
\mathcal{A}_{L U}=-\sqrt{1-\epsilon^2} \frac{\left|\boldsymbol{P}''_{\perp}\right|\left|\boldsymbol{P}'_{ \perp}\right|}{M_N M_2} \frac{\mathcal{C}\left[w_5 \hat{l}_1^{\perp h} D_1\right]}{\mathcal{C}\left[\hat{u}_1 D_1\right]} \sin \Delta \phi\ ,
\end{equation}
where the quark DPDFs $\hat u_1$ and $\hat l_1^{\perp h}$ follow the definition in Refs.~~\cite{Anselmino:2011ss,Anselmino:2011bb} and can be projected through the results of ${\cal M}^{[\gamma^+]}$ and  ${\cal M}^{[\gamma^+\gamma^5]}$ of the last section, respectively.  
In the above equation, $\mathcal{C}$ stands for convolutions of quark DPDFs with the fragmentation function $\mathcal{D}_1$, the $P''$ stands for the momentum of the second hadron probed in the CFR and $M_2$ corresponds to its mass, which is the same as the proton in the case here. Since the process receives contributions from quarks of all flavors, quark DPDFs of all flavors should be summed over, weighted by their charges $\hat{l}_1^{\perp h} = \sum_{q} \bar{e}_q^2~ \hat{l}_{1,q}^{\perp h}$. The $\bar e$ stands for the quark or lepton charge in the unit of $|e|$, i.e., $\bar e_u=2/3$ and $\bar e_d=-1/3$. 

To estimate the spin asymmetry, we further assume that the unpolarized fragmentation contribution from the CFR cancels out from the numerator and denominator and define the following reduced asymmetry,
\begin{align}
\begin{split}
    \mathcal{A}_0 \equiv \frac{\mathcal{C}\left[w_5 \hat{l}_1^{\perp h} D_1\right]}{\mathcal{C}\left[\hat{u}_1 D_1\right]} \approx \frac{\hat{l}_1^{\perp h}(x_B,l_\perp,\xi,t)}{\hat{u}_1(x_B,l_\perp,\xi,t)}\ ,
\end{split}
\end{align}
which depends on the ratio of the structure functions $\hat{l}_1^{\perp h}$ and $\hat{u}_1$ that, again, sum over all quark flavors.

To obtain the quark DPDFs $\hat{l}_1^{\perp h}$ and $\hat{u}_1$, the $x'$ and $x''$ integrals in Eqs.~(\ref{eq:e5},\ref{eq:e6}) 
must be performed. The final results will depend on three form factors: 
\begin{align}
     &\text{Re}\mathcal{H}_C(\xi,t)\equiv \sum_{q} \bar{e}_q^2 \int_{-1}^1 \text{d}x \text{Re} \mathcal {T}_b(x,\xi) H_q(x,\xi,t)\ ,\nonumber\\
     &\text{Im}\mathcal{H}_C(\xi,t)\equiv \sum_{q} \bar{e}_q^2 \int_{-1}^1 \text{d}x \text{Im} \mathcal {T}_b(x,\xi) H_q(x,\xi,t)\ ,\\
     &\text{Im}\mathcal{H}'_C(\xi,x_B,t)\equiv \sum_{q} \bar{e}_q^2 \int_{-1}^1 \text{d}x \text{Im} \mathcal {T}_a(x,\xi,x_B) H_q(x,\xi,t)\ . \nonumber
\end{align}
The hard scattering coefficient $\mathcal {T}_b$ coincides with the one in DVCS~\cite{Ji:1996ek}. Therefore, the corresponding amplitudes will be the ordinary real and imaginary Compton form factors (CFFs) $\text{Re}\mathcal{H}_C(\xi,t)$ and $\text{Im}\mathcal{H}_C(\xi,t)$, whereas the other coefficient $\mathcal {T}_a$ is slightly modified and depends on $x_B$. The corresponding modified CFF $\text{Im}\mathcal{H}'_C(\xi,x_B,t)$ can be explicitly written as,
\begin{align}
\begin{split}
        \text{Im}\mathcal{H}'_C(\xi,x_B,t)=& \pi \sum_{q} \bar{e}_q^2 \Big[H_q((1+\xi)x_B-\xi,\xi,t)\\
        &\mp H_q(-(1+\xi)x_B+\xi,\xi,t)\Big]\ ,
\end{split}
\end{align}
in terms of GPDs. The $\mp$ sign is determined from the parity of the GPDs. For vector GPDs $H$ and $E$ it should be $-$, whereas for axial-vector GPDs $\widetilde{H}$ and $\widetilde{E}$ it should be $+$. Hence, two CFFs and one modified CFF are needed for each quark GPD at each kinematical point $(x_B,\xi,t)$. Here we consider the quark GPDs extracted from the GUMP global analysis of the DVCS measurements as well as lattice GPD calculations~\cite{Guo:2022upw,Guo:2023ahv}. While the extracted CFFs reflect the input DVCS measurements, the modified CFFs are more obscure. Although they resemble the ordinary imaginary CFFs, the modified CFFs are shifted away from the crossover lines at $x=\pm \xi$ due to their extra $x_B$-dependence, making them less constrained by existing experiments. 

\begin{figure}[t]
    \centering
    \includegraphics[width=0.35\textwidth]{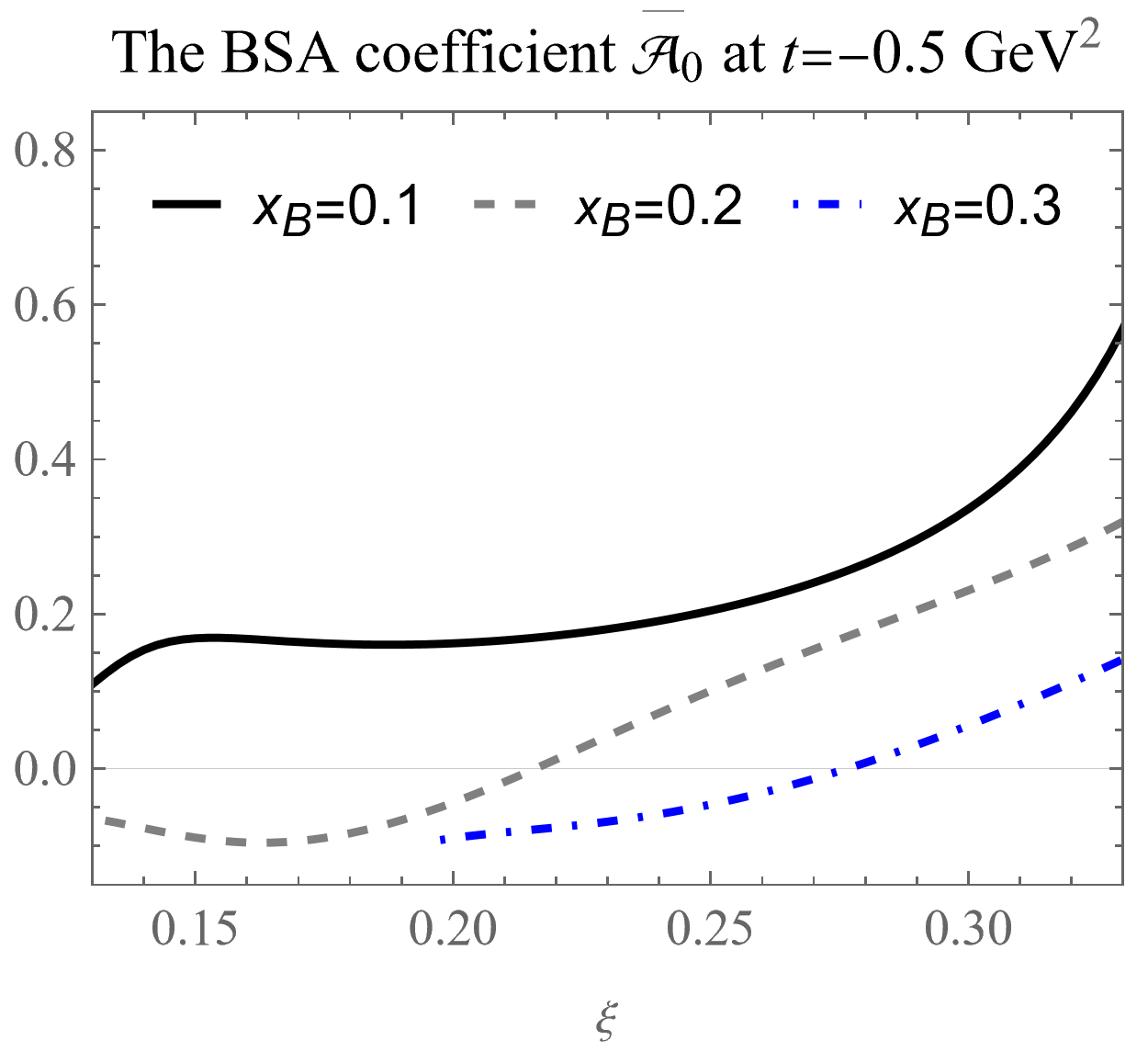}
    \caption{\raggedright A plot of the BSA coefficient $\bar{\mathcal{A}}_0$ at $t=-0.5 \text{ GeV}^2$ and $\mu=2 \text{ GeV}$ with different values of $\xi$ and $x_B$. Since the $\bar{\mathcal{A}}_0$ depends on the modified CFFs, it also depends on the $x_B$ consequently.}
    \label{fig:BSAplt}
\end{figure}

With these (modified) CFFs, the reduced asymmetry $\mathcal{A}_0$ can be evaluated. The perturbative calculation indicates that the $\mathcal{A}_0$ scales with the transverse momentum $l_\perp$ as $l_\perp^{-2}$ for $l_\perp \gg \Lambda_{\rm{QCD}}$. Thus, we define the rescaled coefficient $\bar{\mathcal{A}}_0(x_B,\xi,t)$ according to,
\begin{equation}
    \mathcal{A}_0 = \frac{2 M_N M_2}{l_\perp^2} \bar{\mathcal{A}}_0(x_B,\xi,t)\ ,
\end{equation}
that does not depend on the $l_\perp$. We will present full expressions of the TMD DPDFs and the above asymmetries in a separate publication. Here, we highlight important numeric results from our calculations. In FIG. \ref{fig:BSAplt}, we show the calculated $\bar{\mathcal{A}}_0(x_B,\xi,t)$ at $t=-0.5 \text{ GeV}^2$ and $\mu=2 \text{ GeV}$. The sizable coefficient $\bar{\mathcal{A}}_0(x_B,\xi,t)$ accords with the recent measurement of BSA in this process~\cite{CLAS:2022sqt}.\footnote{The reduced asymmetry $\mathcal{A}_0$ corresponds to the weighted asymmetry therein with an extra minus sign. Except that, the asymmetries here are defined consistently, being aware of the opposite choice of the $z$-direction.} We note that only the dependence of the BSA on the product $P_{T1}P_{T2}$ that corresponds to the product $\Delta_\perp l_\perp $ here has been given~\cite{CLAS:2022sqt}. Therefore, we cannot directly compare the calculated BSA with the measured one without knowing the transverse momentum, which might be soft. Despite that, the observed consistency indicates that the BSA in this process can be interpreted with this semi-exclusive mechanism when final-state interactions are considered. Furthermore, it also implies that such diffractive di-hadron productions or similar diffractive processes can be explored in terms of these exclusive nucleon matrix elements for the nucleon tomography. Although the current work focuses on the region with large transverse momenta, the BSA with soft transverse momenta can be studied in the GTMD framework. This would also allow us to quantitatively compare with the experiments, which will be left to future work.

Finally, we extend the discussion to the single diffractive hadron production in the TFR, where the final-state particles in the CFR are also summed over. Interestingly, the BSA vanishes at the leading twist in this case: since the BSA now corresponds to the azimuthal angular correlation between the leptonic and hadronic planes, it might be in the form of $\epsilon_\perp^{\Delta l^{e}}$, which could be factorized into the leptonic momenta $\boldsymbol{l}^e_\perp,\boldsymbol{l}^{e'}_\perp$ and the hadronic twist-three DPDF $\overline{\mathcal{M}}^{\left[\gamma^\perp\gamma^5\right]}$. The notation $\overline{\mathcal{M}}$ stands for the $l_\perp$-integrated DPDF, distinguished from the unintegrated one. This argument has been examined with explicit calculations of twist-three contributions to the SIDIS in the TFR in Ref.~\cite{Chen:2023wsi} recently, where the twist-three DPDF $\overline{\mathcal{M}}^{\left[\gamma^\perp\gamma^5\right]}$ is shown to contribute to such BSA.

Similar to the twist-two projection, we find the twist-three projection of DPDF $\mathcal{M}^{\left[\gamma^\perp\gamma^5\right]}$ to be non-zero as well, which in fact resembles the twist-two one. The connection between them is analogous to the Wandzura-Wilczek relation for twist-two and twist-three polarized PDFs~\cite{Wandzura:1977qf}, which suggests a corresponding but kinematically suppressed BSA in the case of the single diffractive hadron production in the TFR. However, to obtain the DPDF $\overline{\mathcal{M}}^{\left[\gamma^\perp\gamma^5\right]}$ and make quantitative comparison, the transverse momentum has to be integrated that involves the region with soft transverse momenta inevitably. This can be done in the GTMDs framework, and will be left to future work.

\section{Summary and outlook}
\label{sec:conclusion}

To conclude, we study the SIDDIS process in terms of a semi-exclusive mechanism in this work. We argue that the perturbative QCD can be applied to derive the power behaviors of the quark diffractive PDFs, and calculate the DPDFs in terms of the quark GPDs of the nucleon to the leading order for large transverse momenta. We find non-zero twist-two and twist-three DPDFs $\mathcal{M}^{\left[\gamma^+\gamma^5\right]}$ and $\mathcal{M}^{\left[\gamma^\perp\gamma^5\right]}$ resulting from the final-state interactions, which are responsible for the BSA in the correlated di-hadron productions in the CFR and TFR and the single diffractive hadron production in the TFR, respectively. 

Utilizing the GPDs from global analysis, we look into the BSAs in these processes. For the di-hadron production, the BSA is found to be sizable, in accord with the recent measurement~\cite{CLAS:2022sqt}. Despite the different kinematic regions between the calculation and experiment, the consistency in the BSA implies the potential of these diffractive processes for the study of the nucleon tomography in terms of GPDs and GTMDs. We further discuss the implication of this calculation on the single diffractive hadron production in the TFR, and argue that a similar BSA could exist in this case, which will be kinematically suppressed, though.

For future work, it will be crucial to further investigate the DPDFs with soft transverse momenta in terms of GTMDs for a quantitative comparison of the BSAs to experiments for both the di-hadron and single-hadron productions. Moreover, this work sets an example and motivates one to consider the other observables and processes among all such diffractive processes for the nucleon tomography. As mentioned at the beginning, a comprehensive program to tackle this issue from all possible methods is highly recommended.

\begin{acknowledgments}

We thank Harut Avakian, Yoshitaka Hatta for useful discussions and correspondences. This material is based upon work supported by the U.S. Department of Energy, Office of Science, Office of Nuclear Physics, under contract numbers DE-AC02-05CH11231. The authors also acknowledge partial support by the U.S. Department of Energy, Office of Science, Office of Nuclear Physics under the umbrella of the Quark-Gluon Tomography (QGT) Topical Collaboration with Award DE-SC0023646.

\end{acknowledgments}

\bibliographystyle{apsrev4-1}
\bibliography{refs}
\end{document}